\documentclass{article}
\usepackage{hiph-art}
\usepackage{amsmath,amssymb,epsfig,wrapfig} 

\volnumber{00} \issuenumber{0} \edyear{2004}                             
\frompage{000} \topage{000}                                              
\recrevdate{May 15th, 2004}                                              

\newcommand{\eqeqref}[1]{Eq.~(\ref{#1})}

\newcommand{\refref}[1]{Ref.~\cite{#1}}

\newcommand{\figref}[1]{Fig.~\ref{#1}}

\newcommand{\etal}{{\it et al.}}

\newcommand{\beq}{\begin{equation}}
\newcommand{\eeq}{\end{equation}}
\newcommand{\bea}{\begin{eqnarray}}
\newcommand{\beas}{\begin{eqnarray*}}
\newcommand{\beau}[1]{\begin{equation} \label{#1} \begin{array}{rcl}}
\newcommand{\eea}{\end{eqnarray}}
\newcommand{\eeas}{\end{eqnarray*}}
\newcommand{\eeau}{\end{array} \end{equation}}
\newcommand{\bay}{\begin{array}}
\newcommand{\eay}{\end{array}}
\newcommand{\bals}{\begin{align*}}
\newcommand{\eals}{\end{align*}}

\newcommand{\ds}{\displaystyle}

\newcommand{\ra}{{\rightarrow}}

\newcommand{\vev}[1]{\langle #1 \rangle}

\newcommand{\In}[1]{ \bigg| _{#1} }

\def\k{{\boldsymbol k}}
\def\q{{\boldsymbol q}}
\def\r{{\boldsymbol r}}
\def\x{{\boldsymbol x}}
\def\y{{\boldsymbol y}}

\def\rightmark{Cronin effect from backward to forward rapidity}
\def\leftmark{A.~Accardi}

\title{Cronin effect from backward to forward rapidity: \\ a tale
of two misteries.
}
\authors{ 
{Alberto Accardi %
\index{Accardi, A.} 
\index{}            
}\\[2.812mm]
{\normalsize
Columbia University, Department of Physics\\
538 West 120th Street, New York, NY 10027, USA
}}
 
\abstract{I discuss recent experimental data on the Cronin effect in
  deuteron-gold collisions at the top RHIC energy, in a pseudorapidity range
  $\eta\in[-2,3]$. Two theoretical approaches are compared and
  contrasted: the pQCD-based Glauber-Eikonal model and Colour Glass
  Condensate models. Neither can describe the Cronin
  effect over the whole pseudorapidity interval up to now explored
  experimentally, its most mysterious and intriguing part being at
  negative rapidity.}

 
\begin{document}
 
\maketitle
\setcounter{page}{1}


In hadron-nucleus ($p+A$) and nucleus-nucleus ($A+A$) collisions at
relativistic energy the hadron transverse momentum spectra at moderate
$p_T$$\sim$2-6 GeV are enhanced relative to linear extrapolation from 
$p+p$ reactions. This ``Cronin effect'' has been observed on an energy
range $\sqrt s \approx 20-200$ GeV in both $p+A$ and $A+A$ collisions 
\cite{Cronin}. It is generally
attributed to multiple scattering of projectile partons propagating
through the target nucleus \cite{Accardi02}, and is a sensitive probe
of initial-state modifications of the nuclear wave function ($p+A$
collisions) and of final state in-medium effects ($A+A$ collision) 
\cite{GMcL04}. 
The Cronin effect may be quantified by taking the hadron
$p_T$ spectrum in $p+A$ collision in a given centrality class ($c.cl.$),
normalizing it to binary scaled $p+p$ collisions by the inverse thickness
function $T_A$, and finally dividing it by the $p+p$ spectrum:
\begin{equation}
  R_{pA} =  
    {\frac{1}{T_A(c.cl.)} \frac{dN}{dq_T^2 dy}
       ^{\hspace{-0.3cm}pA\rightarrow h X}
       \hspace{-0.8cm} (c.cl.)} \Bigg/
    {\frac{d\sigma}{dq_T^2 dy}
       ^{\hspace{-0.3cm}pp\rightarrow h X}
       \hspace{-0.8cm} } \ .
  \label{CroninRatio}
\end{equation}
To cancel systematic errors as much as possible, it is also customary
to take the ratio $R_{cp}$ of a given centrality class to the most
peripheral one. 

\section{Recent experimental results at RHIC}
\label{sec:expdata}

\begin{figure}[b]
\begin{center}
\hspace*{-.6cm}
\parbox{14cm}{\epsfig{figure=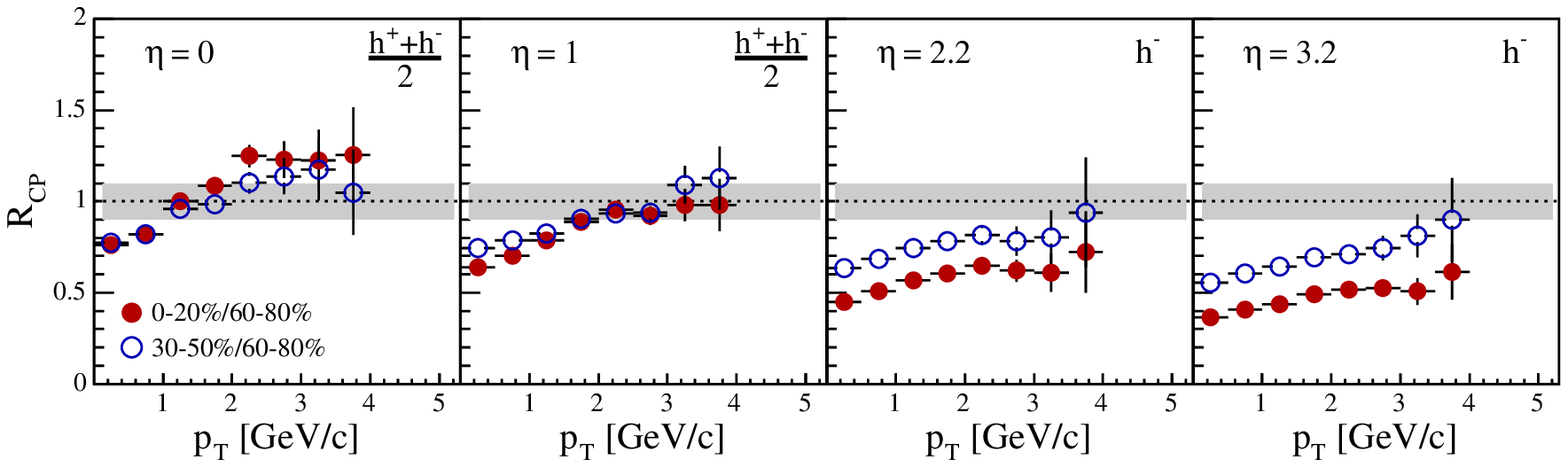,width=14.1cm}}
\hspace*{-.6cm}
\parbox{14cm}{\epsfig{figure=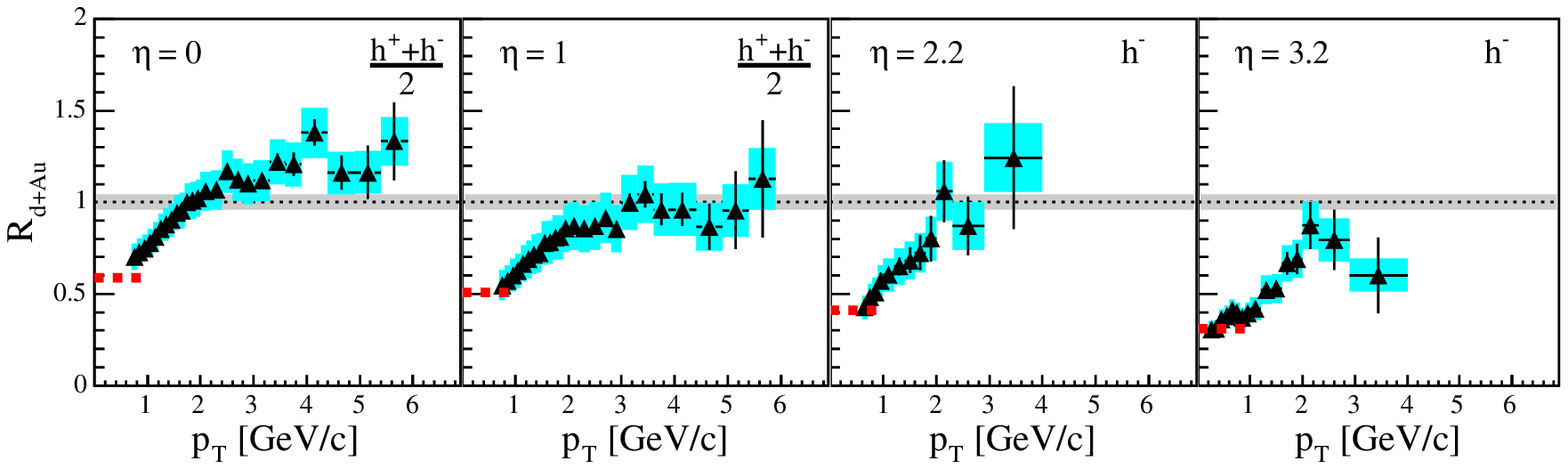,width=14.1cm}}
\vskip-.55cm
\caption{\footnotesize {\bf Top:} $R_{cp}$ measured by BRAHMS \cite{brahmsfwd} at
  rapidity $\eta\in [0,3.2]$ for 2 centrality classes. {\bf Bottom:}
  $R_{dAu}$ in minimum bias collisions, measured
  by BRAHMS \cite{brahmsfwd} at rapidity $\eta\in [0,3.2]$. Note the
  striking difference of the forward rapidity $R_{dAu}$ and $R_{cp}$.}
\label{fig:brahms}.
\end{center}
\vskip-.95cm
\end{figure}

The BRAHMS results \cite{brahmsfwd} on charged hadron $R_{cp}$ in
$d+Au$ collisions at $\sqrt s = 200$ GeV\,A in a rapidity range
$\eta=0-3.2$ are shown in 
\figref{fig:brahms}, top panel. Two 
features are apparent. First, the Cronin peak at $p_T \simeq 3$ GeV in
the mid-rapidity bin is progressively suppressed. At $\eta \gtrsim 2$,
$R_{cp}$ is always smaller than 1, tending to a plateau at larger
transverse momenta. Second, the centrality dependence of the effect is
reversed going from mid- to forward rapidity.

Peripheral collisions are expected to behave similarly to $p+p$
collisions: the more peripheral the collision, the thinner the
nucleus, the smaller the rescattering probability. 
Therefore, the $R_{cp}$ ratio should behave similarly
to $R_{dAu}$. Quite strikingly, this is not the case, as shown in
\figref{fig:brahms}, bottom panel. 
The peak in $R_{dAu}$ is suppressed in going from
$\eta=0$ to $\eta=1$, as confirmed also by 
PHOBOS data \cite{veres}, and looks similar to $R_{cp}$. 
However, at $\eta>2$ the suppression doesn't seem to continue. 
Instead, the data 
grow quite rapidly toward 1, which is reached at $p_T\simeq2$ GeV,
and possibly stay close to 1 at higher $p_T$ though experimental
errors are too large to have a final say. 
This discrepancy is a mystery which cannot be explained up to now by current
theoretical models, and needs further experimental
investigation.

\begin{figure}[t]
\begin{center}
\vskip.0cm
\hspace*{.5cm}
\hspace*{-1cm}\epsfig{figure=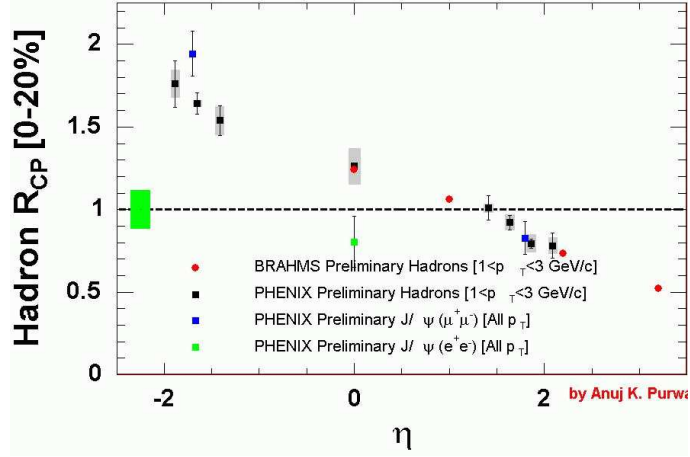,width=9cm}
\vskip-0.5cm
\caption{\footnotesize Summary plot of preliminary results from PHENIX
  and BRAHMS for integrated $R_{cp}$ at $1\le
  p_T \le 3$ GeV. Figure from \refref{Purwar}. }
\label{fig:summary}.
\end{center}
\vskip-0.6cm
\end{figure}

The second and most interesting mystery appears when we 
look at negative rapidities, i.e., at the Au
side. \figref{fig:summary}, compiled by A.Purwar \cite{Purwar}, presents
a summary  plot of the pseudorapidity dependence of the Cronin effect
for charged hadrons integrated on $1<p_T<3$ GeV.
As the rapidity decreases, the gold Bjorken's $x$ increases. 
Then, the parton densities probed by the deuteron decrease and one
would naively expect multiple scatterings and the Cronin effect to
decrease, so that $R_{cp}$ should tend to 1. On the contrary, experimental
data show a steady increase as $\eta$ decreases, reaching $R_{cp}\sim
1.8$ at $\eta=-2$. This large value cannot be accounted
for by standard nuclear effects like anti-shadowing
\cite{EHKS03}, or non perturbative
string fragmentation which pulls hadron production slightly toward
the Au side \cite{Wang03}. Moreover, the same trend at negative $\eta$ appears
also in $J/\psi$ production, ruling out a meson/baryon effect. 

\section{pQCD and Glauber-Eikonal models}
\label{sec:pQCD}

The Glauber-Eikonal (GE) approach \cite{AG03b} to the Cronin
effect treats multiple $2\ra 2$ partonic collisions in collinearly
factorized pQCD, see \figref{fig:GEandCGC}, left panel. 
The low-$p_T$ spectra in $p+A$ collisions are
suppressed by unitarity. At moderate $p_T$, the accumulation  of
transverse momentum  leads to an enhancement of transverse spectra. At
high $p_T$ the binary scaled $p+p$ spectrum is recovered: no
high-$p_T$ shadowing is predicted in this approach.  

The cross-section for production of a parton of flavour $i$, with
transverse momentum $p_T$ and rapidity $y$ in $p+A$ collisions at fixed
impact parameter $b$, is written as 
\begin{eqnarray}
   \frac{d\sigma}{d^2p_T dy d^2b}^{\hspace{-0.5cm}pA\rightarrow iX}
    \hspace{-0.5cm} =
    \vev{xf_{i/p}}_{y_i,p_T} \, \frac{d\sigma^{\,iA}}{d^2p_T dy_i d^2b} 
    \In{y_i=y}
  \hspace*{-.6cm} + \hspace{.1cm}
    T_A(b) \sum_{b}\, \vev{xf_{i/A}}_{y_i,p_T} \, 
    \frac{d\sigma^{\,ip}}{d^2p_T dy_i}  
    \In{y_i=-y} 
 \label{iA}
\end{eqnarray}
where where $T_A(b)$ is the target nucleus thickness function.
Hadron spectra are then obtained as a convolution with the appropriate
fragmentation function $D_{i\ra h}(z,Q)$.

The first term in \eqeqref{iA} accounts for multiple semihard
scatterings of the parton 
$i$ on the target nucleus; in  the second term the nucleus
partons are assumed to undergo a single scattering on the proton.
The average parton flux from the proton, $\vev{xf_{i/p}}$,  
and the parton-nucleon cross-section, $d\sigma^{\,iN}$,
are defined as
\begin{align}
   \vev{xf_{i/p}}_{y_i,p_T} =\,& \ds \frac{K}{\pi} 
      \sum_j \frac{1}{1+\delta_{ij}}  
      \int dy_2 \, x_1f_{i/p}(x_1,Q^2) \,\, 
      \frac{d\hat\sigma}{d\hat t}^{ij}
      x_2f_{j/N}(x_2,Q^2) 
    \Bigg/ 
      \frac{d\sigma^{iN}}{d^2p_T dy_i}
  \nonumber
    \\
  \frac{d\sigma^{iN}}{d^2p_T dy_i} =\,& \frac{K}{\pi}  
    \sum_j \frac{1}{1+\delta_{ij}} 
    \int dy_2 \, \frac{d\hat\sigma}{d\hat t}^{ij}  
    x_2f_{j/N}(x_2,Q^2) 
 \label{avfluxandxsec}
\end{align}
where $\hat t$ is the Mandelstam variable and  $d\hat\sigma/d\hat t$ are
leading order parton-parton cross-sections in collinearly factorized
pQCD. $K$ is a constant factor which takes into account next-to-leading
order corrections.
To regularize the IR divergences of the
single-scattering pQCD parton-nucleon
cross-sections a small mass regulator $p_0$ is introduced in the
propagators, and $Q=\sqrt{p_T^2+p_0^2}/2$ is the scale of the hard process. 
Finally, we introduce a small intrinsic transverse
momentum $\vev{k_T^2}=0.52$ GeV$^2$ to better describe the hadron
spectra in $p+p$ collisions at intermediate $p_T$=1-5 GeV. 
The free parameters $p_0$ and $K$ in Eqs.~(\ref{avfluxandxsec})
are fitted to hadron production data in $p+p$ collisions at the 
energy and rapidity of interest.
This allows to compute 
the spectra in $p+A$ collision and the Cronin ratio with no extra freedom.
For more details, see \refref{AG03b}.

Nuclear effects are included in $d\sigma^{\,iA}$, 
the average transverse momentum distribution of a 
proton parton who suffered at least one semihard scattering:
\begin{align}
   \frac{d\sigma^{\,iA}}{d^2p_Tdyd^2b}
     = \sum_{n=1}^{\infty} \frac{1}{n!} \int & d^2b \, d^2k_1 \cdots d^2k_n
     \, \delta\big(\sum _{i=1,n} {\vec k}_i - {\vec p_T}\big) 
     \nonumber \\
   & \times \frac{d\sigma^{\,iN}}{d^2k_1} T_A(b) 
     \times \dots \times  
     \frac{d\sigma^{\,iN}}{d^2k_n} T_A(b)
     \, e^{\, - \sigma^{\,iN}(p_0) T_A(b)} \,
       \ .
 \label{dWdp}
\end{align}
This equation resums all processes with $n$ multiple $2\ra 2$
parton scatterings. The exponential factor in \eqeqref{dWdp}
represents the probability that the parton suffered no semihard
scatterings after the 
$n$-th one, and explicitly unitarize the cross-section at the
nuclear level. 
Unitarity introduces a suppression of parton yields compared to the
binary scaled $p+p$ case. This is best seen integrating \eqeqref{dWdp}
over the transverse momentum:
$d\sigma^{iA}/{dyd^2b} \approx 1 - e^{\, - \sigma^{\,iN}(p_0)
T_A(b)}$. At low opacity  $\chi = \sigma^{\,iN}(p_0) T_A(b) \ll 1$, 
i.e.,when the number of 
scatterings per parton is small, the binary scaling is
recovered. However, at large opacity, $\chi \gtrsim 1$, the parton yield
is suppressed: $d\sigma^{iA}/{dyd^2b} \ll 1 < \sigma^{\,iN}(p_0) T_A(b)$. This
suppression is what we call ``geometrical
shadowing'', since it is driven purely by the geometry of
the collision through the thickness function $T_A$. As the integrated
yield is dominated by small momentum partons, geometrical shadowing is
dominant at low $p_T$. 
Beside the geometrical quark and gluon shadowing, which is automatically
included in GE models, at low enough $x$ one expects genuine dynamical
shadowing due to non-linear gluon interactions as described in, e.g.,  
Colour Glass Condensate (CGC) models, see Section \ref{sec:CGC}. 

\begin{figure}[t]
\begin{center}
\parbox{14cm}{
\epsfig{figure=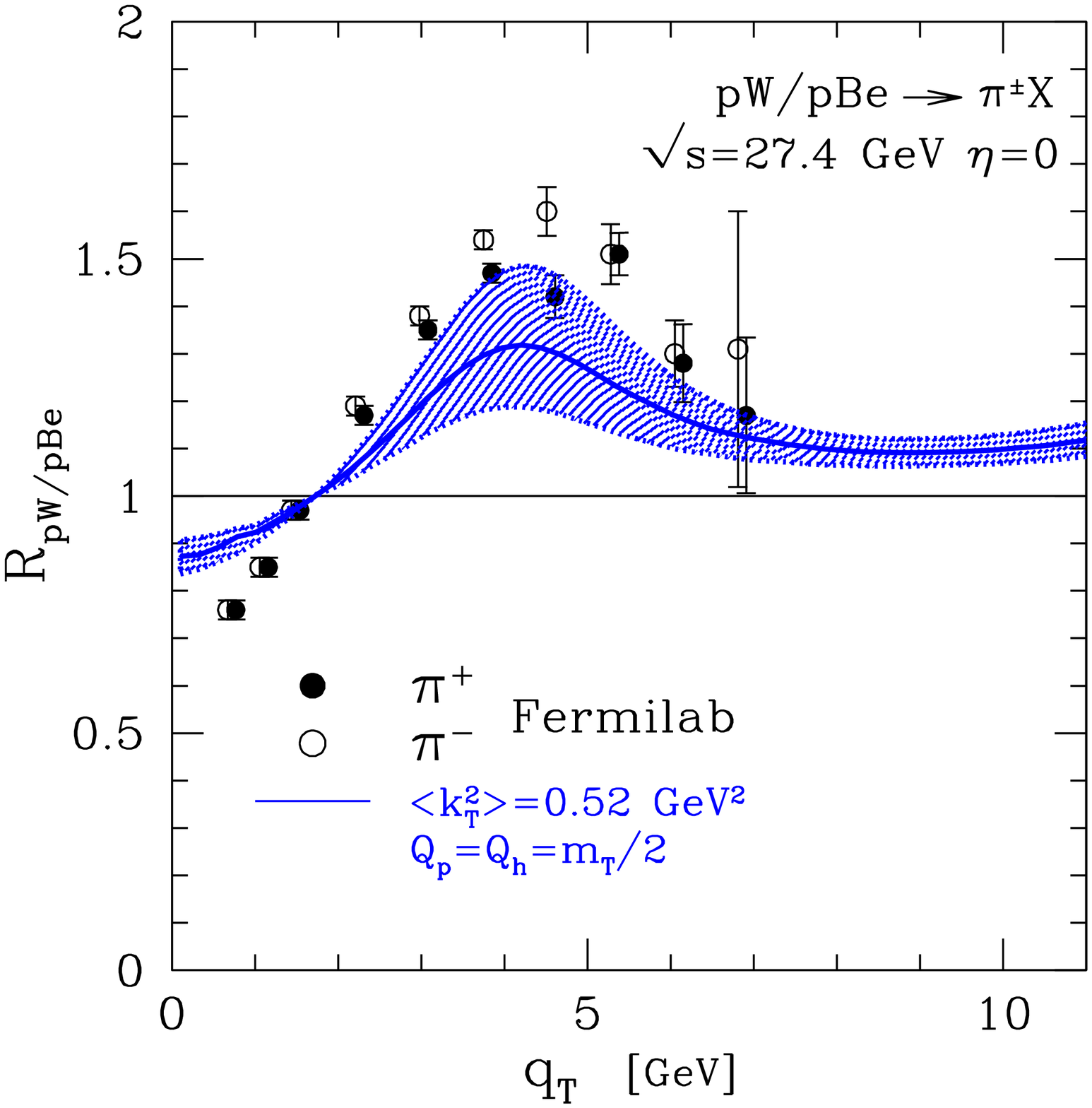,width=4cm}
\epsfig{figure=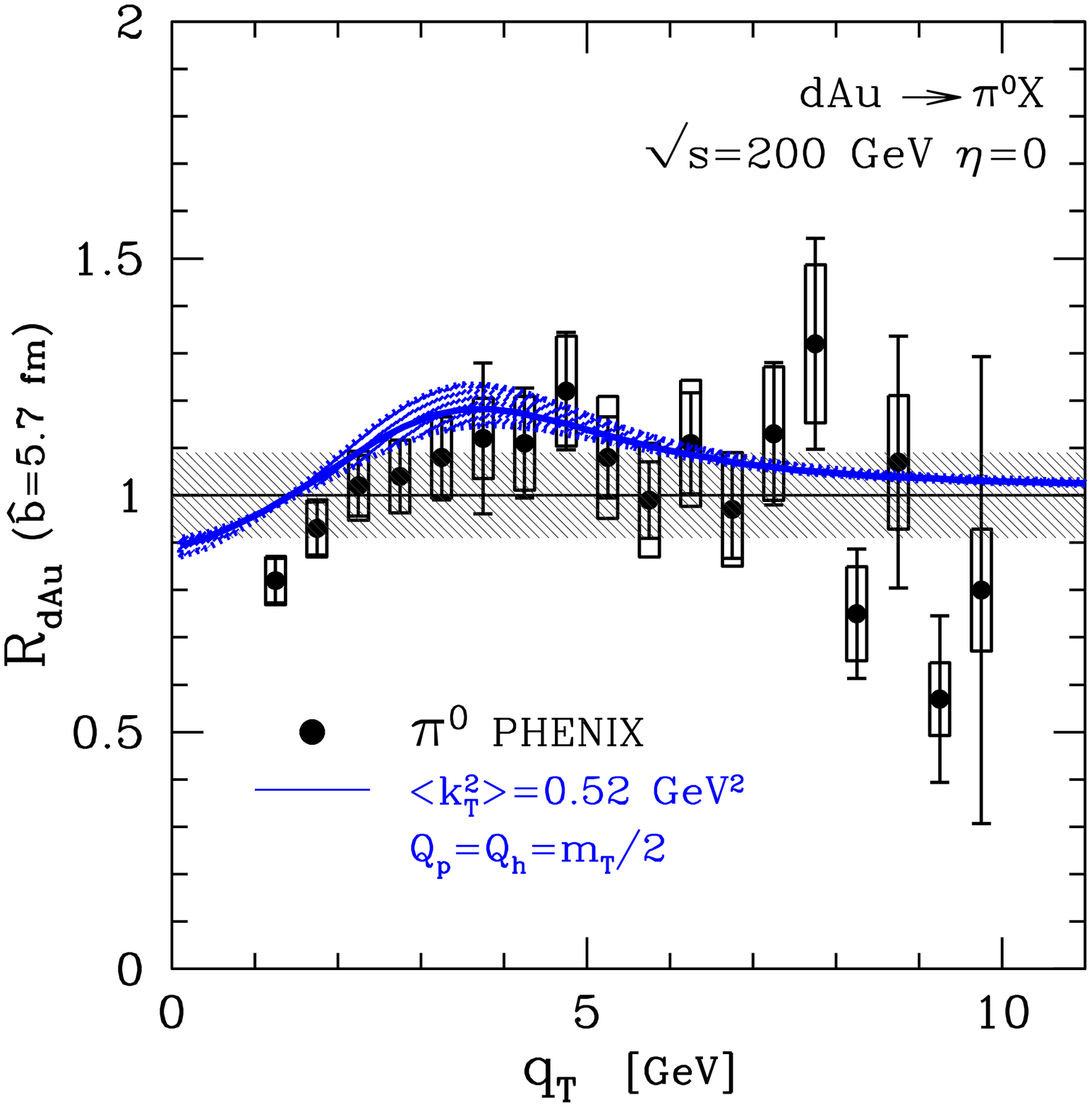,width=4cm}
\epsfig{figure=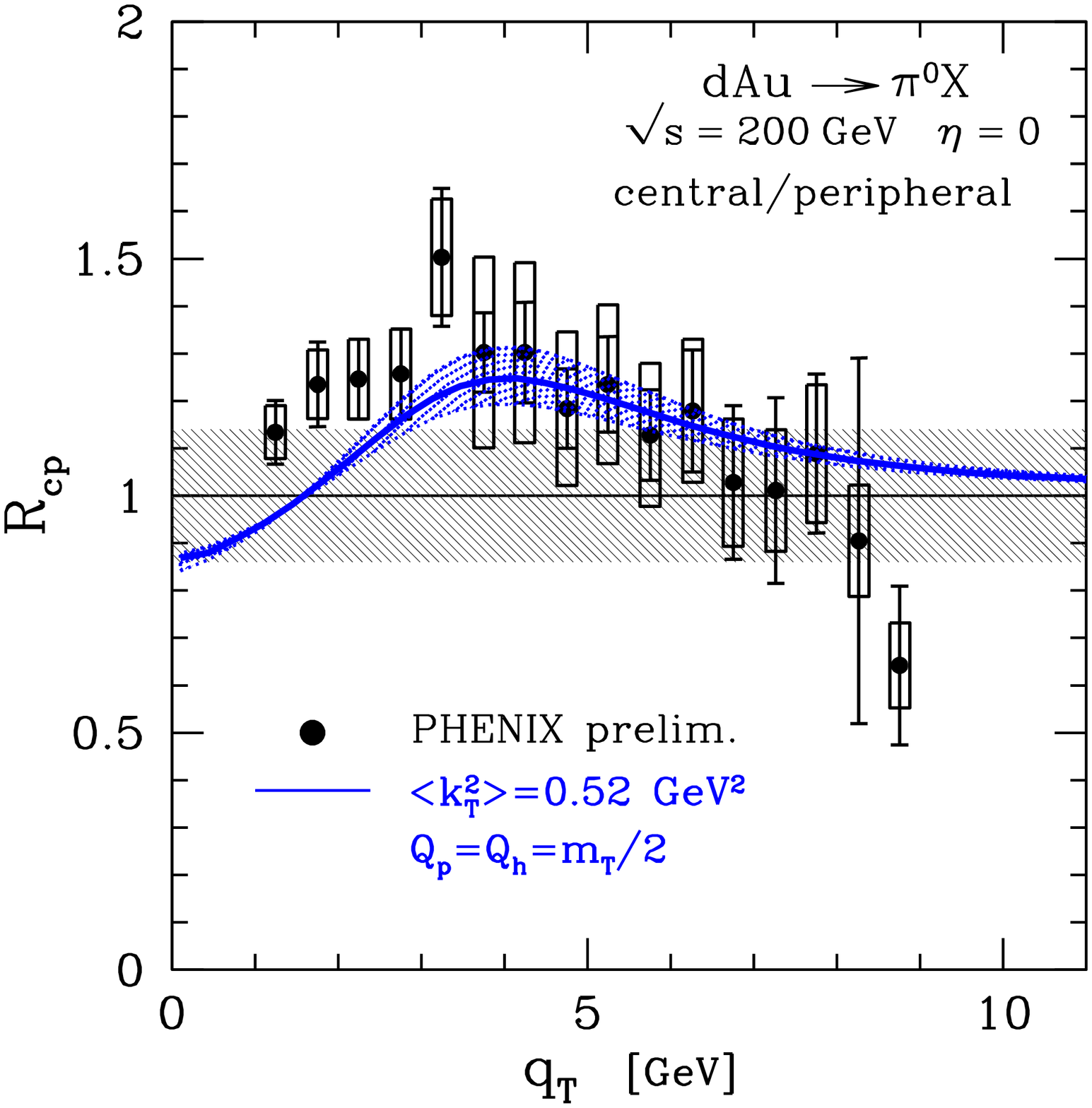,width=4cm}
}
\vskip-.5cm
\caption{\footnotesize Cronin effect on pion production at Fermilab
  \cite{Cronin2} and RHIC \cite{Adler03pA} at $\eta=0$. The solid curve
  is the GE computation. Theoretical errors due to the fit of $p_0$
  are shown as a shaded band around the solid curve. The rightmost panel  
  shows the 0-20\%/60-88\% centrality classes ratio.} 
\label{fig:midrapidity}.
\end{center}
\vskip-.8cm
\end{figure}

The GE model reproduces quite well both
Fermilab and PHENIX data at $\eta=0$ (\figref{fig:midrapidity} left
and middle). It also describe the increase of the Cronin effect with
increasing centrality (\figref{fig:midrapidity}, right). 
The agreement of GE calculations with $y=0$ data, and
especially the centrality dependence, suggest that 
there is no dynamical shadowing nor CGC at RHIC midrapidity. 

To address the BRAHMS data at forward rapidity $\eta\approx3.2$
\cite{brahmsfwd}, we would first need to fit $p_0$ and $K$ in $p+p$
collisions at the same pseudo-rapidity. Unfortunately the available
$p_T$-range $p_T \lesssim 4$ GeV is not large enough for the fit to
be done. Therefore, we use the parameters extracted at $\eta=0$. 
The resulting Cronin ratio, shown by the solid line in Figure 2,
overestimates the data at such low-$p_T$. 
However, the opacity $\chi_0=0.95$ might be
underestimated, due to the use of the mid-rapidity parameters. To
check this we tripled the opacity: the resulting dashed line
touches the data at $p_T\gtrsim 2$ GeV, but predicts a strong Cronin
peak at $p_T\gtrsim 3$ GeV. On the other hand, a similar strong peak
would appear in the computation of $R_{cp}$, in contrast with the data
in \figref{fig:brahms}. 


\section{Colour Glass Condensate models}
\label{sec:CGC}

The Colour Glass Condensate (CGC)
is an effective theory for the nuclear
gluon field at small-$x$ \cite{Iancu02}.
The valence quarks, treated as a collection of random colour sources
$\rho$, radiate the gluon field $A$. At low-$x$ the gluon occupation
number is so large that the 
gluon field can be treated semi-classically and computed as a solution
of Yang-Mills equation of motion in the presence of colour sources.
The theory is characterized by a saturation scale $Q_s$. Gluons with
momenta lower than this scale are in the ``saturation'' regime: their
density is so high that gluon-gluon fusion processes limit a further
growth. At large momenta the gluon field is in the perturbative
``parton gas'' regime. A recently conjectured
intermediate ``geometric scaling window'' may extend at momenta
$Q_s<p_T<Q_s^2/Q_0$, where quantum effects from the saturation region
further modify the evolution from the perturbative to the saturation
region. 

\begin{figure}[t]
\begin{center}
\parbox{5cm}{\epsfig{figure=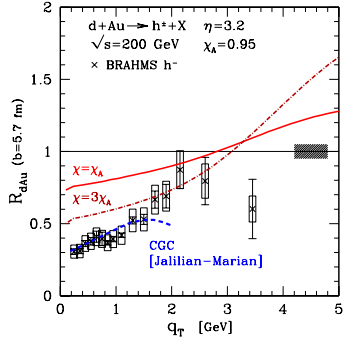,width=4.5cm}}
\begin{minipage}{6cm}{\caption{Cronin ratio at $\eta=3.2$. 
    The solid and dot-dashed
    curves are the GE model computation with $\chi=\chi_0$ and
    $\chi=3\chi_0$, respectively. The dashed curve is a CGC
    computation by Jalilian-Marian \cite{JJ04}. Data are taken from
    \cite{brahmsfwd}. \label{fig:fwdrap} }}
\end{minipage}
\end{center}
\vskip-.5cm
\end{figure}

Observables ${\cal O}[A[\rho]]$ are computed as an average over the
colour sources with a weight $W_y[\rho]$ depending on the gluon
rapidity $y=\log(1/x)$ as follows:
$
  \vev{{\cal O}[A]}_y = \int D\rho \, W_y[\rho]  {\cal O}[A[\rho]]  
$. 
Gluons at a given $x$ are themselves colour sources for gluons at
smaller $x'<x$. This evolution of the gluon field with $x$ is
captured by the so-called JIMWLK evolution equation \cite{Iancu02}. 

If one approximates the proton as a dilute colour
source, gluon production in $p+A$ collisions can be explicitly 
written in a $k_T$-factorized form \cite{BGV04-I}:
\begin{equation*}
\frac{d\overline{N}_g}{d^2\q_\perp dy}
=\frac{1}{16\pi^3 q_\perp^2}
\int \frac{d^2\k_\perp}{(2\pi)^2}  \varphi_p(\q_\perp-\k_\perp)
\Big[ k_\perp^2 
\int d^2\r_\perp e^{i\k_\perp\cdot\x_\perp} 
\left<
U^\dagger(0)
U({\x_\perp})
\right> \Big] \ .
\end{equation*}
Here, the proton's unintegrated wave function $\varphi_p$ is
multiplied by the squared scattering amplitude on the target nucleus,
expressed as a correlator of two Wilson lines. 
When the weight $W$ is Gaussian, 
\begin{equation*}
  W_y[\rho] = {\cal N}_y \exp \Big[ \frac12 \int_{-\infty}^y dy
    \int d\x_\perp d\y_\perp 
    \frac{\rho^a(\x_\perp)\rho^a(\y_\perp)}{\lambda_y(x_\perp-y_\perp)}
    \Big] \ , 
\end{equation*}
gluon production in $p+A$ collision can be interpreted 
as multiple $2 \ra 1$ partonic scatterings as illustrated in
\figref{fig:GEandCGC}, center.  

\begin{figure}[t]
\begin{center}
\epsfig{figure=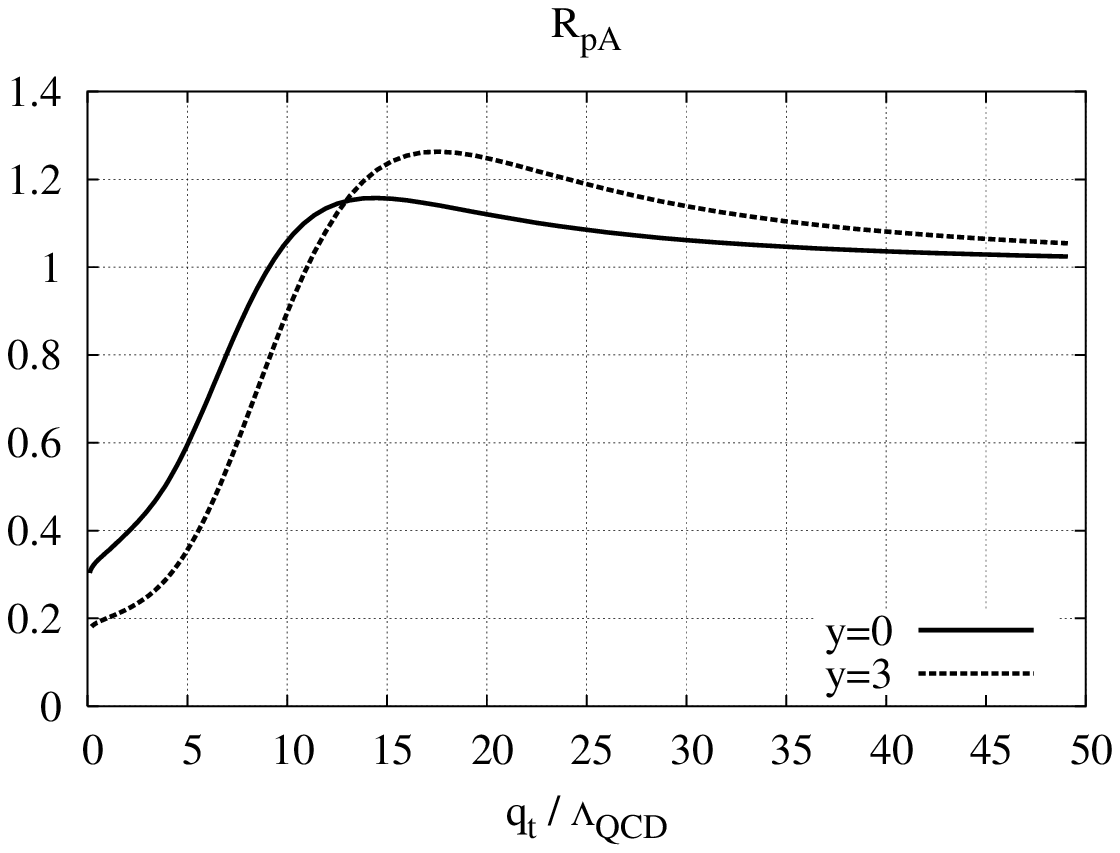,width=5.1cm}
\hspace*{.5cm}
\epsfig{figure=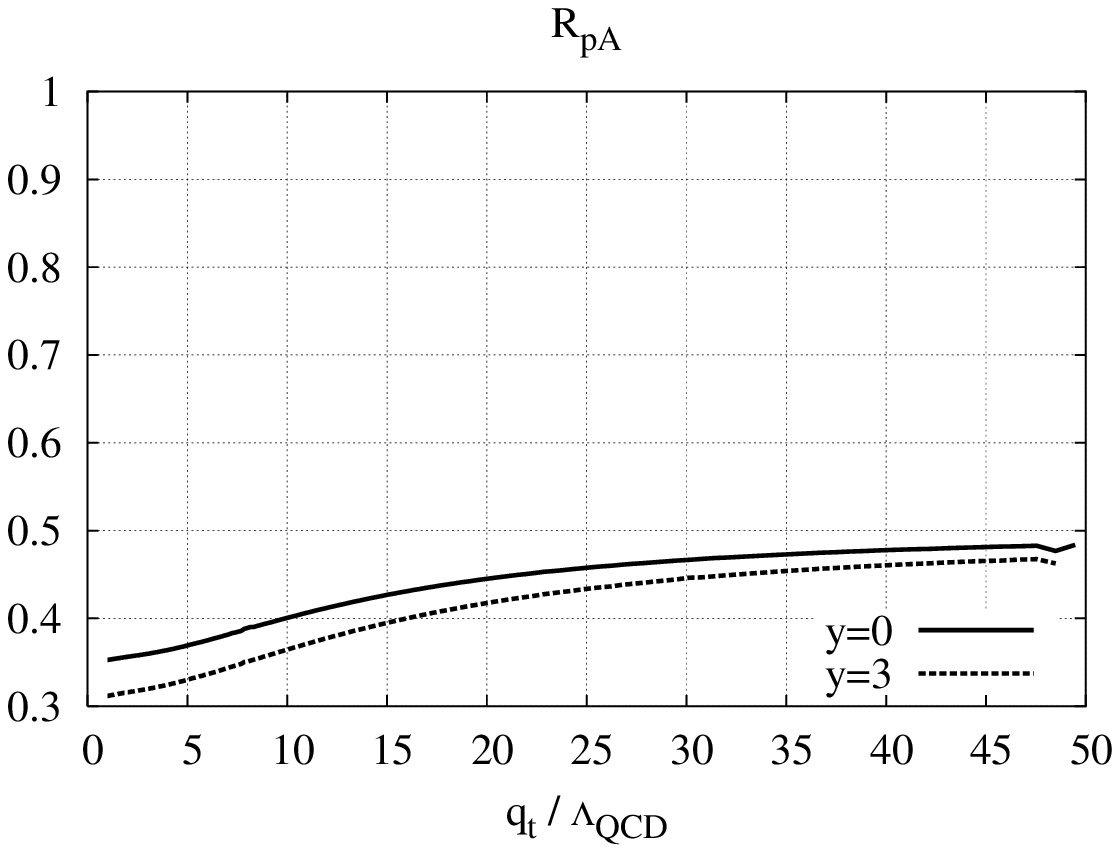,width=5.1cm}
\vskip-.6cm
\caption{\footnotesize Cronin effect in the MV model with naive
  quantum evolution (left) and with full evolution in the ``deep
  saturation'' limit (right). Figure taken from \refref{BGV04-I}.} 
\label{fig:CGC}.
\end{center}
\vskip-.8cm
\end{figure}

To proceed further, one needs to choose a specific model for the
colour sources. The McLerran-Venugopalan model assumes the gluon
correlations to be local: $\lambda(\x_\perp - \y_\perp) =
Q_{s}(x) \delta(\x_\perp - \y_\perp)$ where $Q_{s}(x) = 
\alpha_s^2 \frac{8\pi N_c}{N_c-1} xG(x,Q^2)$. Quantum evolution of the
gluon field is included naively in the $x$-dependence of $Q_{s}$ only. 
An estimate from \refref{BGV04-I} of the Cronin effect on gluon
production is presented in \figref{fig:CGC}, left. It
qualitatively agrees with experimental data at $y=0$. However, at
$y=3$ it predicts  a
rightward shift and an increase of the Cronin peak (analogously to the
GE model). It is disfavored by the $R_{cp}$ data, 
though not incompatible with $R_{dAu}$.  

On the other hand, at larger rapidities (i.e., at small $x$) the
system undergoes a long quantum evolution, and enters the saturation
regime. Thus, we should solve the JIMWLK equation. 
In the Gaussian approximation, a self-consistent solution in the
saturation regime exists, but with non-local gluon
correlations \cite{IIM02}. In other words, the gluon density has
become so high that gluons are no more bound to single nucleons, but
correlate over macroscopic distances.  
The resulting Cronin ratio is highly suppressed over the whole $p_T$
range, in qualitative agreement with $R_{cp}(y=3)$. However, it
disagrees both with the forward $R_{dAu}(y=3)$ and with mid-rapidity
data. 

A word of caution: at large $\eta$ valence quark
production is important, if not dominant, as demonstrated by the
excess of positively charged hadrons over negatively charged ones in
$p+p$ collisions at
$\eta=3.2$ anf $p_T>2$ GeV \cite{Debbe04}.
Therefore, these computations should be considered
only as illustrative of the physics included in the CGC approach.
However, even the more quantitative computation of
\refref{JJ04}, who takes into account valence quarks and their  
fragmentation into charged hadrons, fails to reproduce the $p_T>1.5$
GeV of $R_{dAu}$ at $y=3$, see \figref{fig:fwdrap}\footnote{A very
  recent computation \cite{KKT04b} was more successful. 
  It is based on a parametrization of the onset and rapidity
  dependence of quantum evolution which mimics CGC effects, 
  and models valence quark scatterings in a different
  way than in \refref{JJ04}.}.

\section{Where does pQCD meet the CGC?}
\label{GEandCGC}

\begin{figure}[t]
\begin{center}
\parbox{12.4cm}{
\parbox{8cm}{\epsfig{figure=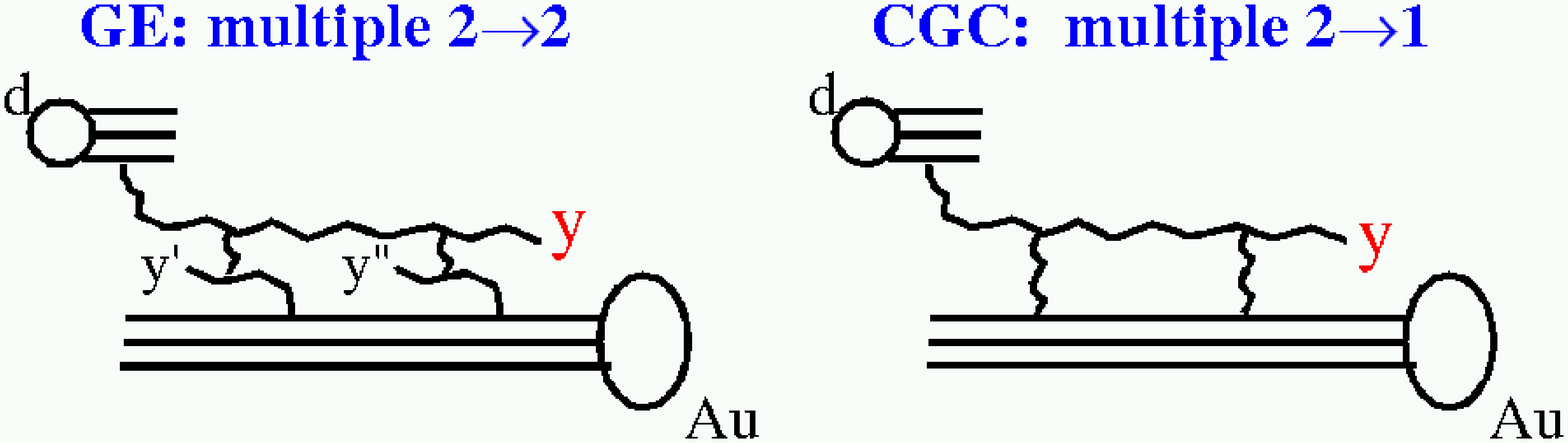,width=7.8cm}}
\parbox{4cm}{\epsfig{figure=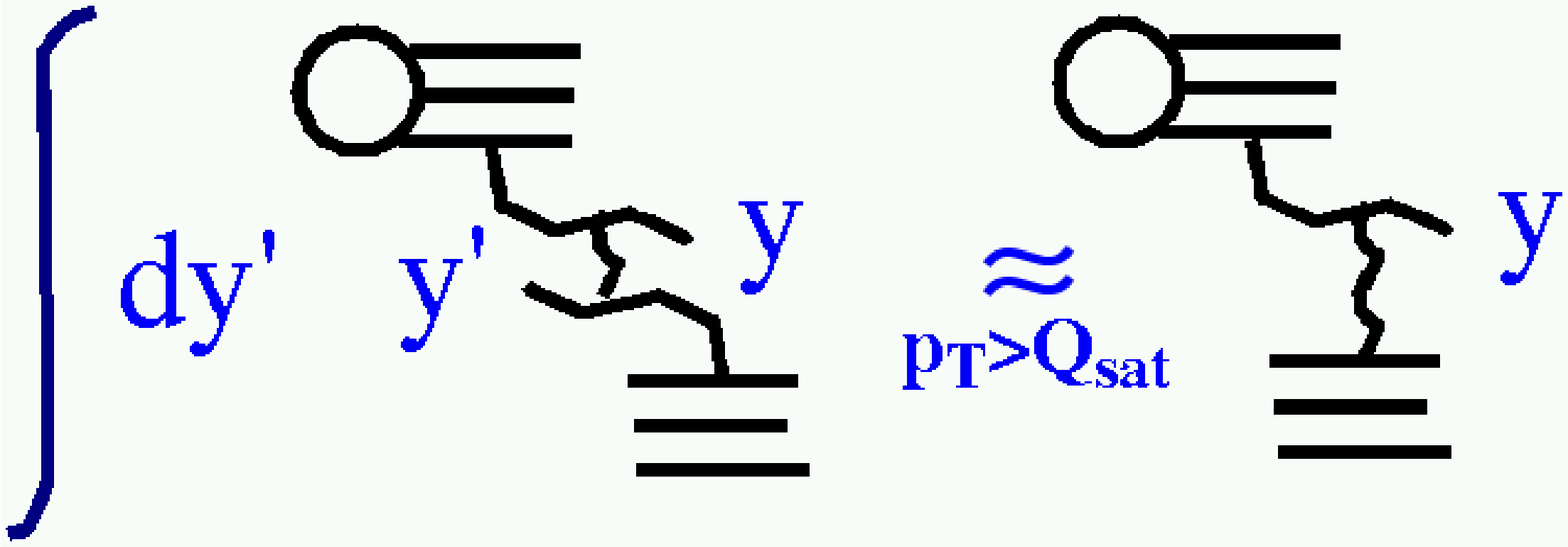,width=4.2cm}}
}
\vskip-.6cm
\caption{\footnotesize {\it Left and middle:} Multiple scattering
  processes included in the GE and CGC models, respectively. {\it
  Right:} Illustration of the equivalence of pQCD and the MV model at
  high $p_T$ discussed in \refref{GMcL97}.} 
\label{fig:GEandCGC}.
\end{center}
\vskip-.8cm
\end{figure}

At mid-rapidity, GE and CGC models give similar results for the Cronin
effect. Does this mean that  to some extent they are equivalent, 
even if GE models describe multiple $2\ra 2$ scatterings and CGC
multiple $2\ra1$ scatterings? 
The standard answer is yes. Consider the single scattering terms
depicted in \figref{fig:GEandCGC}, right. The single-inclusive
cross section in pQCD is obtained by integration over the unobserved
parton rapidity $y'$. In the limit of 
large $p_T\gg Q_s$ and asymptotic energy $p_T/\sqrt s e^{-y} \ll 1$,     
one obtains in the MV model  $d\sigma_{pQCD} \approx d\sigma_{pQCD}
\propto 1/p_T^4$ \cite{GMcL97}. 
This is usually read as a proof that the single-inclusive cross section
in the MV model reduces to pQCD in the high-$p_T$ limit. Hence the GE
and the MV models are equivalent. 

\begin{figure}[t]
\begin{center}
\vskip-.3cm
\parbox{7.3cm}{
\hspace*{-0.5cm}
\epsfig{figure=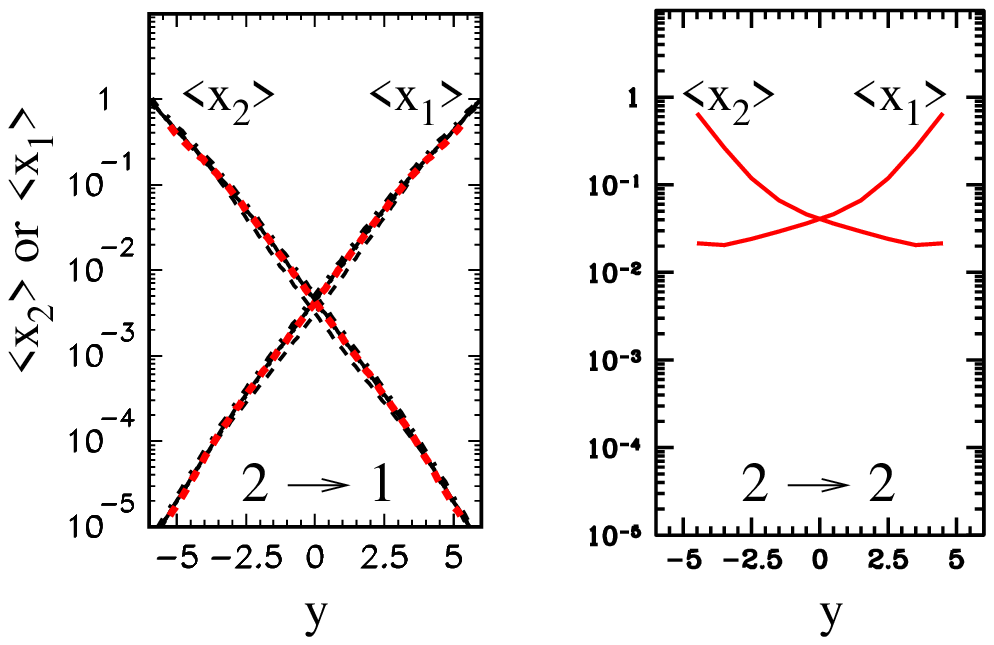,width=8.6cm}
}
\begin{minipage}{5.2cm}{\caption{\footnotesize Average $\vev{x_1}$ and
      $\vev{x_2}$ in $p+p$ collisions at RHIC energy, for $2\ra 1$
      processes (left, taken from \refref{Szczurek03}) and $2\ra 2$
      processes (right). } 
  \label{fig:avx} }
\end{minipage}
\end{center}
\vskip-1.1cm
\end{figure}

However, the above equivalence can hold only in a restricted kinematic
domain. This is clearly seen if we compute the average Bjorken's $x$'s 
probed in $p+p$ collisions. In the $2\ra 2$ kinematics, they depend on the
rapidity of both the observed and the unobserved parton:
\begin{equation*}
  x_{1,2}^{2\ra 2} = \frac{p_T}{\sqrt s} \big( e^{\pm y} + e^{\pm y'} \big) \ .
\end{equation*} 
In $2\ra 1$ parton processes, the observed particle fixes them completely:
\begin{equation*}
  x_{1,2}^{2\ra 1} = \frac{p_T}{\sqrt s} \big( e^{\pm y} \big) \ .
\end{equation*}
As shown in \figref{fig:avx}, the $2\ra 1$ and $2\ra 2$ processes
probe the target at quite different Bjorken's $x$!
The reason of this difference 
is that in the $2\ra 2$ processes one has to integrate over
the unobserved particle, over a quite large phase space $\Delta
y'\approx 10$. When $y$ is large, 
most of these particles are produced at small or
negative rapidity, which compensate the decrease in $\vev{x_2}$ expected when
the rapidity of the observed particle increases.

Given \figref{fig:avx}, one is tempted to ask ``Which of the two models
is wrong: GE or CGC?'' In my opinion, this is the wrong question.
The existence of $2\ra 2$ hard scattering processes at mid-rapidity
has been experimentally proved by the observation of back-to-back
hadrons at large $p_T$ in $p+p$ and $p+A$ collisions \cite{STARbtb}.
On the other hand, there is no reason to
doubt about the existence of $2\ra 1$ processes. 
So a more pertinent question is ``What is the interplay
between $2\ra 2$ and $2\ra 1$ processes? Is there a region where one
is dominant over the other?'' 

We may have a clue at the answer from the following
zeroth-order argument. Neglecting the quarks for ease of notation, the
single inclusive cross-section for $2\ra 2$ processes may evaluated as:
\begin{align*} 
  \frac{d\sigma^{pp \ra gX}}{dy} & =
    \int dy' \, x_1 G\big(x_1\big) \, x_2 G\big(x_2\big) \,
    \frac{d\hat\sigma}{d\hat t} \\
    & \approx \,  
    {\Delta y'} \times xG\big({\vev{x_1}_{2\ra2}}\big) \, 
    xG\big({\vev{x_2}_{2\ra2}}\big) \,
    \frac{d\hat\sigma}{d\hat t}
\end{align*}
where $\Delta y'$ is the width of the integration interval over
$y'$, and $\vev{x_{1,2}}_{2\ra2}$ can be read off \figref{fig:avx}.
In the $2\ra 1$ case we have:
\begin{align*} 
  \frac{d\sigma^{pp \ra gX}}{dy} 
    \,\approx \, xG\big({\vev{x_1}_{2\ra1}}\big) \, 
    xG\big(\vev{x_2}_{2\ra1}\big) \,
    {d\hat\sigma\over d\hat t}
\end{align*}
where the value of $\vev{x_{1,2}}_{2\ra1}$ is shown in
\figref{fig:avx}. At mid-rapidity, $\vev{x_2}_{2\ra 1} \approx  
\vev{x_2}_{2\ra 2}$, then $2\ra 2$ processes are dominant because of a
larger phase-space ($\Delta y' \approx 10$). At very forward rapidity, 
$\vev{x_2}_{2\ra 1} \ll  \vev{x_2}_{2\ra 2}$; hence, $2\ra 1$ processes
are dominant because of the small-$x$ growth of parton distributions
(the same argument can be used to analyze backward rapidities).

The transition between these two regimes is poorly known even in $p+p$
collisions, and should be more carefully studied, both theoretically
and experimentally. This requires a formalism
which is able to describe both processes in a unified
framework. Collinear factorized pQCD is not suited, as kinematics
forbids $2\ra 1$ processes if incoming partons have zero transverse
momentum. On the other hand, $k_T$ factorization can describe $2\ra 1$
processes, and the possible production of a second or more hard
particles is hidden in the unintegrated distribution functions for the
incoming partons \cite{smallx,Szczurek03}. 
It is thus very suited for the problem at hand.

\section{Conclusions}
\label{sec:conclusions}

The pQCD based GE model is able to quantitatively describe the Cronin
effect at midrapidty over a broad $\sqrt{s}=20$-$200$ GeV energy range,
indicating absence of CGC.
However, at large rapidity it predicts an increasing Cronin effect, 
which is unfavoured by experimental data. Only
incoherent $2\ra 2$ multiple parton scatterings are included in the
GE model; adding coherence effects would suppress the Cronin peak, but
still predict $R_{dAu} \sim R_{cp} \sim 1$ at large $p_T$ \cite{QV04}.
On the other hand, CGC models describe $2\ra 1$ multiple coherent
parton scatterings, and suggest at RHIC a transition from a
non saturated region at mid-rapidity, to a saturated region at forward
rapidity. Here, nonlocal gluon correlations created during quantum
evolution suppress below 1 the otherwise increasing Cronin ratio.
This predicted plateau at high-$p_T$ is the hallmark of CGC.
Unfortunately, the discrepancy between $R_{dAu}$ and $R_{cp}$ BRAHMS
data, which is not present in any of the discussed models, 
does not allow to conclude whether this picture is correct or not. 

At mid-rapidity GE and CGC models are usually thought to be equivalent, 
even though they describe different partonic subprocesses, and probe
quite different Bjorken $x$'s at forward and backward rapidity.
Simple kinematic considerations suggest a dominance of $2\ra 2$
processes at mid-rapidity, and of $2\ra 1$ processes at forward
and backward rapidity. 
The interplay and transition between the two is poorly known, and 
needs to be studied more accurately already at the $p+p$ level, 
then extended to $p+A$ collisions. The appropriate tools are 
already on the market: $k_T$-factorized pQCD on the theory side, and  
the STAR and PHENIX detectors on the experimental side. Azimuthal
two-hadron correlations and their rapidity dependence 
are an especially suited observable for this purpose.

The steep increase of $R_{cp}$ at negative rapidity adds a new dimension
to the problem. Indeed both GE and CGC models expect the Cronin effect
to be more and more reduced at $\eta<0$. This is due to the decrease 
of the nucleus opacity $\chi_A(y)$ in one case, and of the saturation
scale $Q_s(y)\propto e^y$ in the other. Standard nuclear
effects like anti-shadowing are too weak to explain the magnitude of
the negative rapidity Cronin effect, which requires a new theoretical
explanation: either a neglected effect in the GE and CGC models, or
a new piece of physics. \\[-.2cm]


{\bf Acknowledgements.} I would like to thank M.Gyulassy, P.Levai,
A.Purwar, J.W.Qiu, A.Szczurek and X.N.Wang for inspiring discussions.
\\[-.7cm]

\end{document}